%% file: main.tex
\documentclass[conference]{IEEEtran}
\IEEEoverridecommandlockouts

\usepackage{cite}
\usepackage{amsmath,amssymb,amsfonts}
\usepackage{graphicx}
\usepackage{textcomp}
\usepackage{fancyhdr}
\usepackage[hyphens]{url}
\usepackage{hyperref}
\usepackage{xcolor}
\usepackage{algorithm}
\usepackage[noend]{algpseudocode}
\usepackage{subcaption}
\usepackage[utf8]{inputenc}
\usepackage[noabbrev]{cleveref}
\usepackage{mathptmx} 
\def\BibTeX{{\rm B\kern-.05em{\sc i\kern-.025em b}\kern-.08em
    T\kern-.1667em\lower.7ex\hbox{E}\kern-.125emX}}

\providecommand{\linesref}[2]{\hyperref[#1]{Lines~\ref*{#1}--\ref*{#2}}}
\providecommand{\lineref}[1]{\hyperref[#1]{Line~\ref*{#1}}}

\def\algbackskip{\hskip\dimexpr-\algorithmicindent+\labelsep}
\def\LState{\State \algbackskip}%

\begin{document}

\title{Fast and Scalable Sparse Triangular Solver for Multi-GPU Based HPC Architectures}

\author{\IEEEauthorblockN{Chenhao Xie\IEEEauthorrefmark{1}, Jieyang Chen\IEEEauthorrefmark{2}, Jesun S Firoz\IEEEauthorrefmark{1}, Jiajia Li\IEEEauthorrefmark{1}, Shuaiwen Leon Song\IEEEauthorrefmark{3}, Kevin Barker\IEEEauthorrefmark{1}, Mark Raugas\IEEEauthorrefmark{1}, Ang Li\IEEEauthorrefmark{1}}
	\IEEEauthorblockA{\IEEEauthorrefmark{1}Pacific Northwest National Laboratory (PNNL), Richland, USA\\
	\IEEEauthorrefmark{2}Oak Ridge National Laboratory, Oak Ridge, USA\\
	\IEEEauthorrefmark{3}Future System Architecture (FSA) Lab, University of Sydney, Sydney, Australia\\
		\IEEEauthorrefmark{1}(chenhao.xie, jesun.firoz, jiajia.li, kevin.barker, mark.raugas, ang.li)@pnnl.gov, \\
		\IEEEauthorrefmark{2} chenj3@ornl.gov,
		\IEEEauthorrefmark{3}shuaiwen.song@sydney.edu.au}
}

\maketitle


\begin{abstract}

Designing efficient and scalable sparse linear algebra kernels on modern multi-GPU based HPC systems is a daunting task due to significant irregular memory references and workload imbalance across the GPUs. This is particularly the case for \textit{Sparse Triangular Solver (SpTRSV)} which introduces additional two-dimensional computation dependencies among subsequent computation steps. Dependency information is exchanged and shared among GPUs, thus warrant for efficient memory allocation, data partitioning, and workload distribution as well as fine-grained communication and synchronization support. In this work, we demonstrate that directly adopting unified memory can 
adversely affect the performance of SpTRSV on multi-GPU architectures, despite linking via fast interconnect like NVLinks and NVSwitches. Alternatively, we employ the latest \textit{NVSHMEM} technology based on \textit{Partitioned Global Address Space} programming model to enable efficient fine-grained communication and drastic synchronization overhead reduction. 
Furthermore, to handle workload imbalance, we propose a malleable task-pool execution model which can further enhance the utilization of GPUs. 
By applying these techniques, our experiments on the NVIDIA multi-GPU supernode V100-DGX-1 and DGX-2 systems demonstrate that our design can achieve on average 3.53$\times$ (up to 9.86$\times$) speedup on a DGX-1 system and 3.66 $\times$ (up to 9.64$\times$) speedup on a DGX-2 system with 4-GPUs over the Unified-Memory design. The comprehensive sensitivity and scalability studies also show that the proposed zero-copy SpTRSV is able to fully utilize the computing and communication resources of the multi-GPU system.

\end{abstract}

\begin{IEEEkeywords}
Sparse Linear Algebra Kernels, Triangular Solver, Multi-GPU Systems, Task Model
\end{IEEEkeywords}

\input{sec-introduction.tex}
\input{sec-background.tex}
\input{sec-character.tex}

\input{sec-shmem.tex}
\input{sec-taskmodel.tex}

\input{sec-evaluation.tex}

\input{sec-conclusion.tex}



\bibliographystyle{ieeetr}
\bibliography{main}

\end{document}

%% file: sec-introduction.tex
\section{Introduction}

Multiple Graphics Processing Units (GPUs) based server architectures such as NVIDIA DGX-1\&2\cite{dgx1} become increasingly popular in modern HPC systems for further enhancing the acceleration of high throughput workloads. 
Due to the ever-increasing size of data, the demand of out-of-memory execution is ever-increasing in emerging domains including deep learning, big data analytics, and planet-scale applications.  
Since the recent GPUs on multi-GPU systems have their own local memory and distributed memory space, executing large-scale sparse linear algebra operations on these multi-GPU systems presents several challenges, especially for the case of \textit{Sparse Triangular Solver (SpTRSV)} kernel in which strong inner-dependency among the computation steps is inherent. However, the existing optimization 
techniques (e.g., synchronization-free execution\cite{liu2017fast,jiyaCapellini}) for sparse linear algebra kernels either focus on multi-CPUs system where dependency can be handled by sequentially execution \cite{wang2018} or assume that the problem data always fits into a single GPU\cite{naumov2011parallel,liu2017fast,jiyaCapellini}. They do not offer orchestrated strategies to improve the scalability in multi-GPU platforms.


SpTRSV is considered as the main kernel for many important HPC applications, such as numerical simulation, power grid simulation and optimization, computational chemistry, and climate modeling, in the direct solution of linear systems and least squares problems, structured-grid problems, the preconditioners of iterative methods, etc \cite{vuduc2002automatic,vuduc2003automatic,mayer2009parallel, wolf2010factors,wang2018,Sao2019,acceleration2016,ding2020leveraging,SpTRSV2018grid,powergrid}.

Given $Lx=b$ or $Ux = b$, where $L$ is the \underline{L}ower triangular matrix, $U$ is the \underline{U}pper triangular matrix, $b$ is the vector of values, SpTRSV seeks the solution for vector $x$. We refer to an entry of $x$ as a component. Solving a component may require certain dependencies to be met. 
In this paper, we consider designing SpTRSV on a single compute node with multiple GPUs (i.e., \emph{scaling-up}). We partition the larger matrices among the GPUs. Hence, to calculate the dependencies for each component, GPUs may require to share information with each other. Unless proper consideration is given, such inter-GPU information exchange can introduce \textit{severe inter-GPU data movement and synchronization problems}. Additionally, due to the partitioning strategy of columns of $L$ matrix and $b$ in the baseline distribution model, components are solved in the ascending 
order (\Cref{sec:back}). 
This can adversely affect workload balance due to uneven waiting time of the GPUs which, in turn, can degrade the  performance of SpTRSV on a multi-GPU system.


To tackle the challenges above, we propose an efficient and scalable design for the parallel execution of Sparse Triangular Solver on modern multi-GPU systems.   

First, to facilitate easy data sharing and memory allocation, the state-of-the-art method involves using the unified memory introduced by the major vendors (e.g., NVIDIA), which spans across all the processors (CPU and GPUs) in the system. However, we find that enabling the Unified Memory for efficient data exchange is often counterproductive (\Cref{sec:uni-mem}). 
To improve this, we employ the latest \textit{NVSHMEM} \cite{potluri2016simplifying} technology based on the \textit{OpenSHMEM} \cite{openshmem_website,potluri2017openshmem} specification. NVSHMEM provides high-level application programming interface for partitioned global address space. It supports fine-grained one-sided \emph{get} and \emph{put} operations, and peer-to-peer (P2P) communication directly from GPU kernels which result in opportunities to overlap computation and communication. With NVSHMEM, we propose a `\emph{zero-copy}' mechanism to eliminate page migration and avoid memory contention. In our proposed design, the dependencies of SpTRSV are accumulated on each GPU's symmetric heap and asynchronously communicated using the \emph{get} operation. Experiments demonstrate that our NVSHMEM-based design overcomes the limitations of the Unified Memory approach and improves the performance of SpTRSV significantly.

Second, to tackle workload imbalance on GPUs, we propose a task distribution model that not only distributes component more evenly among GPUs but also ensures that the independent calculations are scheduled on different GPUs, thus ensuring high utilization of the GPU hardware. We finally demonstrate that a combination of fine-grained communication, efficient overlapping between computation and communication, synchronization-free execution, and thoughtfully designed load balancing helps achieve better overall performance and scalability. In summary, this paper makes the following contributions:
\begin{itemize}
    \item Through performance characterization, we identify that applying the state-of-the-art Unified Memory for data sharing among GPUs may cause severe performance penalty (\Cref{sec:uni-mem}).
    \item We leverage the new \textit{NVSHMEM} technology to design an efficient and scalable algorithm for the SpTRSV kernel executing on a multi-GPU system setup. It has profound design implication for a spectrum of applications that have inherent irregular memory accesses and strong inner-task dependencies (\Cref{sec:nvshmem}).
    \item For better workload balancing, a novel task scheduling scheme is further introduced (\Cref{sec:task_model}).
    \item We demonstrate the performance benefit and scalability of our proposed SpTRSV design on a range of inputs that require out-of-core execution (\Cref{sec:evaluation}).
\end{itemize}

%% file: sec-background.tex
\section{Background}
\label{sec:back}


\subsection{Dependencies of SpTRSV}

The \emph{sparse triangular solver operation} (SpTRSV) is an important building block of many numerical linear algebra applications, which is to solve the following linear equations:
\begin{equation}
    Lx = b \quad \quad \text{or} \quad\quad Ux = b
\end{equation}
where $L$ and $U$ are square lower and upper triangular matrices, $x$ and $b$ are the solution and right-hand-side vectors. Based on the processing order, solving $Lx=b$ is often referred to as \emph{forward substitution} while solving $Ux=b$ is referred to as \emph{backward substitution}.

\begin{algorithm} \small
	\caption{Forward substitution triangular solver algorithm for $Lx=b$}\label{alg:Serial}
		\begin{algorithmic}[1]
		    \Statex $\textbf{Input: } L_{n\times n}\text{, } b_{n}$
		    \Statex $\textbf{Output: } x_{n}$
			\Procedure{Lower Triangular Solver(Lx=b)}{}
			\State {Allocate size-$n$ left.sum array and $0$-initialized.} \label{line:lft_sum_decl}
			\For {$i = 0 \text{ to }  n-1$}\label{line:iter_x}
		    \State $ x_i \gets (b_i - \text{left.sum}_i)/l_{ii} $
		    \For {$j = i+1 \text{ to } n-1$}
		    \State $ \text{left.sum}_j \gets \text{left.sum}_j + l_{ij} * x_i$ \label{line:partial_left_sum}
			\EndFor
			\EndFor
		
			\EndProcedure
		\end{algorithmic}	
\end{algorithm}

In comparison with other sparse basic linear algebra subprograms (BLAS) \cite{duff2002overview} such as sparse matrix-vector multiplication (SpMV) and sparse-sparse matrix-matrix multiplication (SpGEMM), SpTRSV is more challenging to parallelize due to the inherent complex dependencies within the linear system. Since backward substitution follows the similar procedure as forward substitution (i.e., solving $x$ in descending order), in this work, without loss of generality, we use forward substitution for illustration.

\begin{figure}[t]
	\centering
	\begin{minipage}[h]{0.23\textwidth}
		
		\includegraphics[width=1\textwidth]{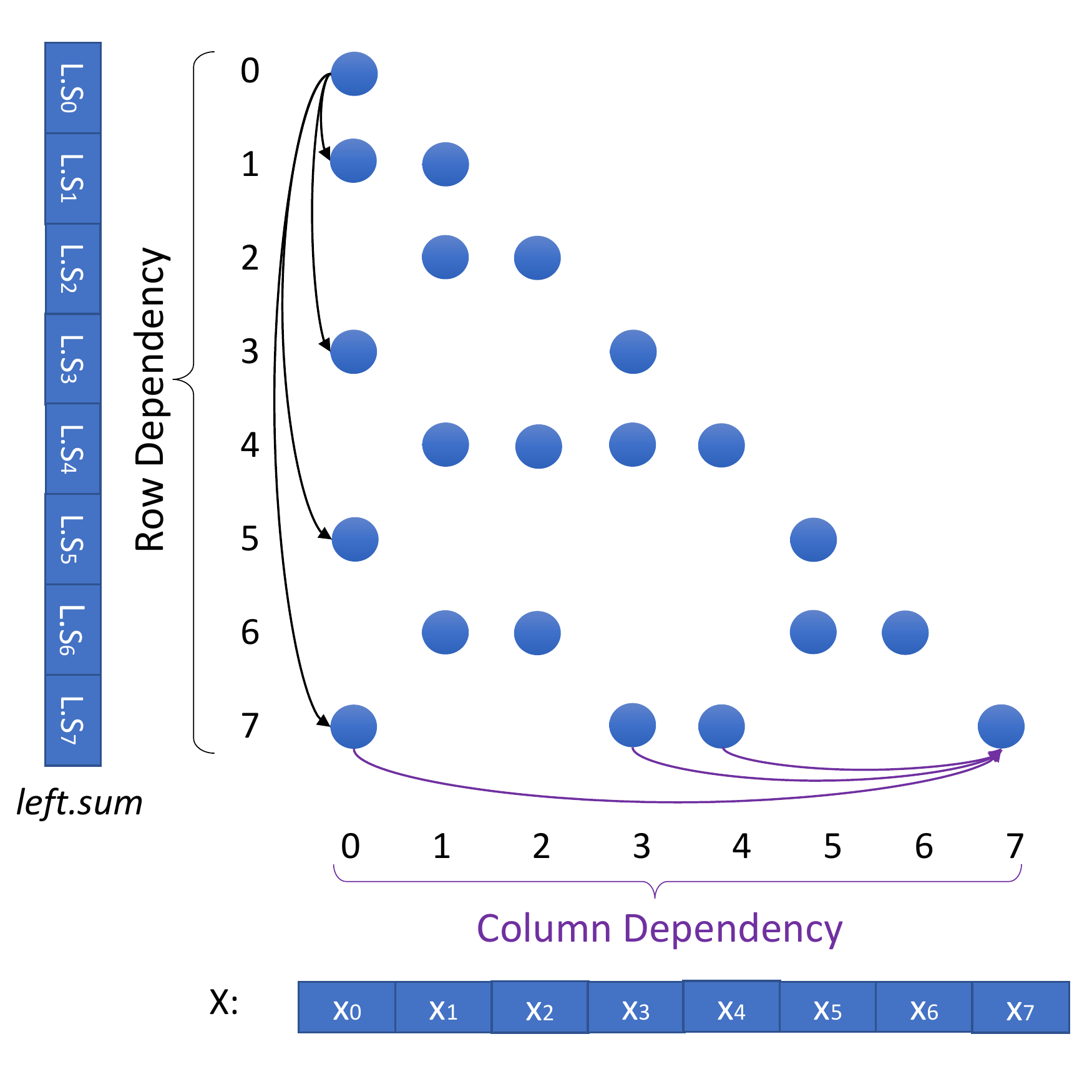}
		\vspace{-0.6cm}
		\subcaption{Dependencies in L's matrix}\label{fig:Lmatrix}
	\end{minipage}
	\begin{minipage}[h]{0.23\textwidth}
		
		\includegraphics[width=1\textwidth]{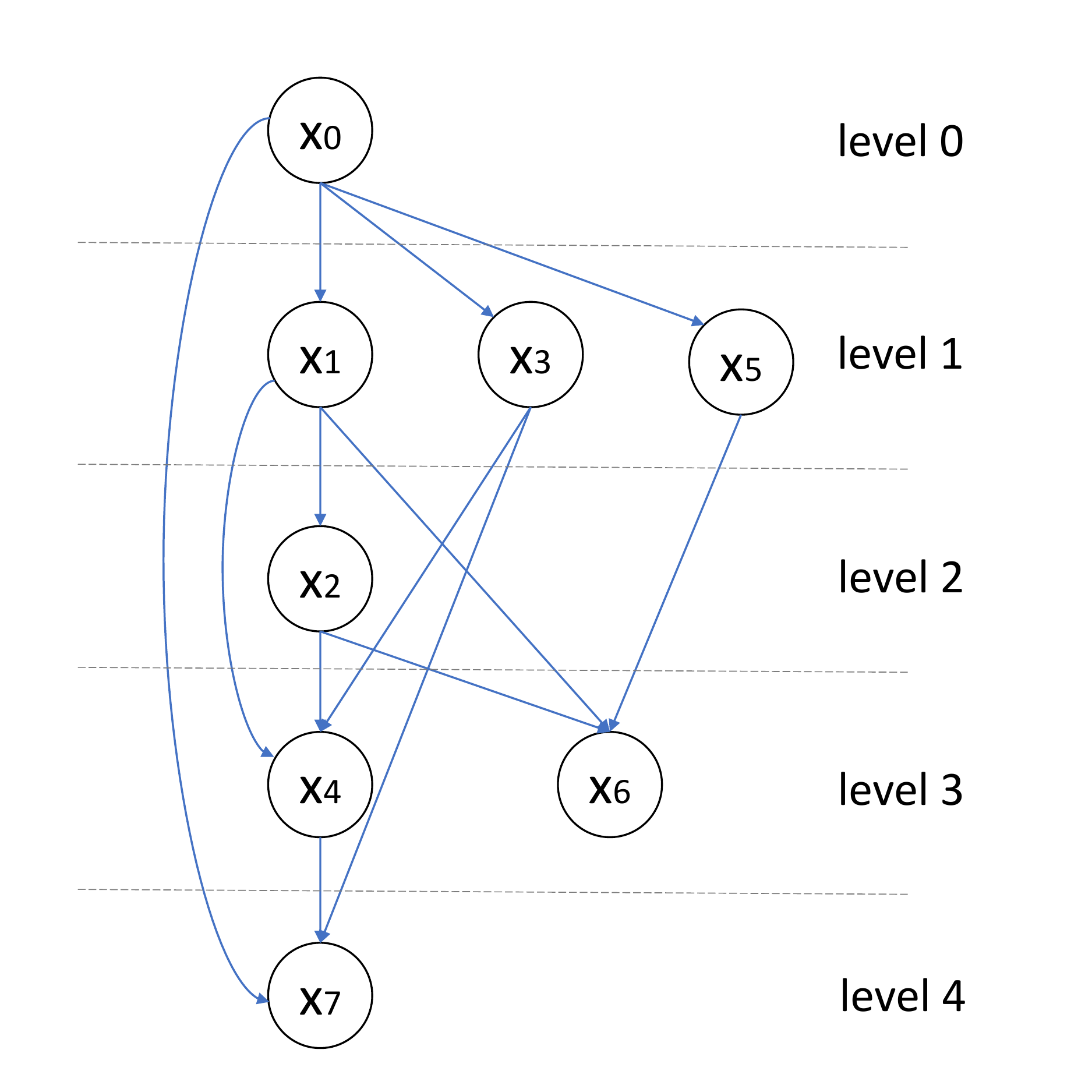}
		\vspace{-0.6cm}
		\subcaption{Level-sets of $x_i$}\label{fig:level-sets}
	\end{minipage}
	\caption{The dependencies in a lower triangular and parallel SpTRSV using level-set method.}
		\vspace{-0.5cm}
	\label{fig:L_level_dep}
\end{figure}


\Cref{alg:Serial} shows a typical serial implementation for solving {a dense} $Lx = b$. The algorithm accesses all columns in ascending order (\lineref{line:iter_x}) to solve each component of the solution vector $x$ using an intermediate array $left.sum$ (\lineref{line:lft_sum_decl}). $left.sum$ maintains the partial sums computed during the intermediate steps corresponding to the nonzero entries of the columns in $L$. In each step, the value of one component, $x_i$, is calculated and then  $left.sum_j$ values are updated by computing $x_i*l_{ij}$ and adding this value to $left.sum_j$ (\lineref{line:partial_left_sum}). In the subsequent iterations, these $left.sum_j$ values are used to compute $x_{i+1}$ components.

Two types of dependencies in the algorithm make it challenging to parallelize the SpTRSV kernel. First, all $l_{ij}$ in the same row {$i$} need to be accessed to update $left.sum_i$, which creates \textit{column dependency}. Second, the $left.sum_j$ depends on $x_i$ and $l_{ij}$, which causes \textit{row dependency}. 
\Cref{fig:L_level_dep}(a) shows an example of $L$ matrix and the column and row dependencies for the non-zero values (nnz) at the last row and the first column, respectively. In order to calculate a component $x_7$, components $x_0$, $x_3$ and $x_4$ have to be calculated due to the column dependency while all of $left.sum_1$, $left.sum_3$, $left.sum_5$ and $left.sum_7$ is row dependent on $x_0$. 

If we assume the input matrix $L$ is stored in the compressed sparse column (CSC) format, even if the zero-valued entries of $l_{ij}$ can be skipped, the row and column dependencies still exist. 



\subsection{Parallel SpTRSV}\label{subsec:related_work}



The parallel solutions of sparse triangular linear systems have been studied by many researchers \cite{naumov2011parallel,wang2018,totoni2014structure,anzt2015iterative,Sao2019,park2014Sparsifying,kabir2015stsk,bradley2016hybrid,dufrechou2018level,liu2017fast}.
The common strategy comes from the observation that some components are independent and can be processed simultaneously. For example, in the dependence graph of \Cref{fig:Lmatrix}, components $x_1$, $x_3$ and $x_5$ only depend on component $x_0$. Therefore, they can be processed in parallel after $x_0$. By analyzing the $L$ matrix firstly, the components can be reorganized into a number of \emph{level-sets} where the components within a set can be solved in parallel as shown in \Cref{fig:Lmatrix}.
Naumov~\cite{naumov2011parallel} first proposed the single GPU version of sparse triangular solver based on this strategy using CUDA's parallel programming paradigm. 
However, the performance of the parallel algorithm based on level-sets can be limited for several reasons, especially when the matrices exhibit higher number of dependencies among different components. First, the number of components within each set may be small which reversely lead to reduced utilization of the system due to low workload on GPUs. 
Second, there can be a very large number of levels generated from the complex dependencies among the components. In this case, synchronization after each level is mandatory, which can be problematic. Finally, the analysis phase of generating the dependency graph can incur substantial preprocessing overhead,  degrading the efficiency of parallel execution if the solver phase is only invoked a few times.    

\subsection{Synchronization-Free SpTRSV} \label{subsec:syn-free}
To address the inefficiency of level-based parallel SpTRSV algorithm on GPUs, Liu et al. \cite{liu2017fast} propose a fast synchronization-free algorithm for parallel SpTRSV on GPUs. The basic idea is to activate all components on GPUs at the beginning. However, each component will not be processed until the dependencies are satisfied. To track how many dependencies are currently outstanding, a simple pre-processing step calculates the in-degrees (dependencies) for each component. Each of these components are separated into two phases: \emph{lock-wait} and \emph{solve-update}. In the lock-wait phase, a component waits in the loop until the column dependency is satisfied. After that, the component is unlocked to solve the value and then update its dependants based on the row dependence. Rather than the expensive prepossessing step that involves dependency graph computation and synchronization, the components are scheduled by the hardware scheduler of the GPU. Compared with the level-based parallel SpTRSV, synchronization-free mechanism can immediately solve the component as long as all the dependencies are met. 
	

%% file: sec-character.tex
\section{SpTRSV with Unified Memory}
\label{sec:uni-mem}

\begin{algorithm}\small
	\caption{SpTRSV on multiple GPUs with unified memory}\label{alg:Unified}
		\begin{algorithmic}[1]
		    \Statex $\textbf{Input: } col.ptr_{n} \text{,  } row.idx_{nnz} \text{,  } val_{nnz} \text{,  } B_{n}$
		    \Statex $\textbf{Output: } X_{n}$
		    
			\Procedure{SpTRSV-Unified}{}
			
			\State $\text{cudaMallocManaged(*s.left.sum, n)}$\label{line:alloc_left_s}
			\State $\text{cudaMallocManaged(*s.in.degree, n)}$
			\State $\text{cudaMemset (*s.left.sum, 0)}$
			\State $\text{cudaMemset (*s.in.degree, 0)}$\label{line:assignval}
		    \LState $\textit{Get in.degree for all x:}$
		    \ForAll {$d \in ngpu$} \Comment{parallel in GPUs}\label{line:calc_indegree_start}
			\ForAll {$i \in nnz$} \Comment{parallel in threads}
		    \State	$\text{s.atomic.incr(\&s.in.degree[row.idx[i]]}$ \label{line:calc_indegree_end}
			\EndFor
			\EndFor
			\LState $\textit{Solve x:}$
			
			\ForAll {$d \in ngpu$} \Comment{parallel in GPUs}
			\State $\text{cudaMalloc(*d.left.sum[d], dev\_n[d])}$\label{line:device_alloc_d}
			\State $\text{cudaMalloc(*d.in.degree[d], dev\_n[d])}$
			\State $\text{cudaMemset(*d.left.sum[d], 0)}$ 
			\State $\text{cudaMemset(*d.in.degree[d], 0)}$\label{line:device_assign_d}
			\ForAll {$i \in n$} \Comment{parallel in warps}
			\While {$\text{d.in.degree[d][i]+1} \neq \text{s.in.degree[i]}$ } \label{line:lock_start}
			\State $\text{Lock()}$
			\EndWhile \label{line:lock_end}
			\State $\text{x[i]} \gets \text{b[i] - d.left.sum[d][i] - s.left.sum[i]}$\label{line:solve_start}
			\State $\text{x[i]} \gets \text{x[i]/val[col.ptr[i]]}$\label{line:solve_end}
			\ForAll {$j \in nnz_i $} \Comment{parallel in threads}\label{line:update_start}
			\State $ \text{rid} \gets \text{row.idx[j]}$
			\If {$rid \in dev\_d$}\label{line:choose_device}
			\State $\text{d.atomic.add(\&d.left.sum[d][rid], val[j]x[i])}$
			\State $\text{d.atomic.incr(\&d.in.degree[d][rid])}$
			\Else \label{line:choose_uva}
			\State $\text{s.atomic.add(\&s.left.sum[rid], val[j]x[i]))}$
			\State $\text{s.atomic.decr(\&s.in.degree[rid])}$\label{line:update_end}
			\EndIf
			\EndFor
			\EndFor
			\EndFor
			
			\EndProcedure
		\end{algorithmic}	
\end{algorithm}



\subsection{Communication Through Unified Memory}
The main objective of this work is to design an efficient parallel SpTRSV algorithm on a multi-GPU system. To enable inter-communication among the GPUs to exchange the dependency information, we first attempt to leverage CUDA's \emph{Unified Memory} programming model on modern GPUs. Unified Memory provides a single memory address space, where a pool of managed memory is accessible from both CPUs and GPUs using a single pointer within a multi-GPU system\cite{li2015evaluation,
yu2019quantitative}. One of the most salient feature of Unified Memory is that the system automatically migrates data allocated in Unified Memory (using \emph{cudaMallocManaged} API) between the host and device.

Inspired by the aforementioned work \cite{liu2017fast} (\Cref{subsec:syn-free}), our algorithm leverages synchronization-free execution mechanism to handle the dependencies among the components and eliminates the cost of system-level barrier synchronization. \Cref{alg:Unified} shows the pseudocode for our multi-GPU SpTRSV algorithm with Unified Memory. We assume that each solution component $x$, the related right-hand-side vectors $b$ and the columns of matrix $L$  are evenly distributed on the multi-GPU system. Matrix $L$ is represented in a compressed sparse column (CSC) format consisting of three arrays: \textit{col.ptr}, \textit{row.idx}, and \textit{val}. 
We maintain two intermediate arrays, \textit{in.degree} and \textit{left.sum} of size $n$, to keep track of the number of unfinished dependencies and the partial sums calculated from previously calculated components respectively.
Compared to aforementioned work\cite{liu2017fast}, the distributed data post several unique algorithmic design distinctions discussed below.

To keep track of the dependencies and enable inter-communication between GPUs, our algorithm allocates the intermediate arrays in the unified memory space (\textit{s.left.sum} and \textit{s.in.degree} in \linesref{line:alloc_left_s}{line:assignval}) to enable data sharing and remote inter-communication. In addition, for device-local updates and faster access during intermediate calculation for dependencies, the algorithm also allocates device-local arrays (\textit{d.left.sum} and \textit{d.in.degree} in \linesref{line:device_alloc_d}{line:device_assign_d}).
By distinguishing between the device-local updates and unified memory updates 
the algorithm reduces unnecessary remote data access between two GPUs which has been shown to be a major obstacle for 
performance scaling on multi-GPU systems~\cite{arunkumar2017mcm,milic2017beyond}.

\begin{figure}[t]
	\centering
	\begin{minipage}[h]{0.46\textwidth}
		
		\includegraphics[width=1\textwidth]{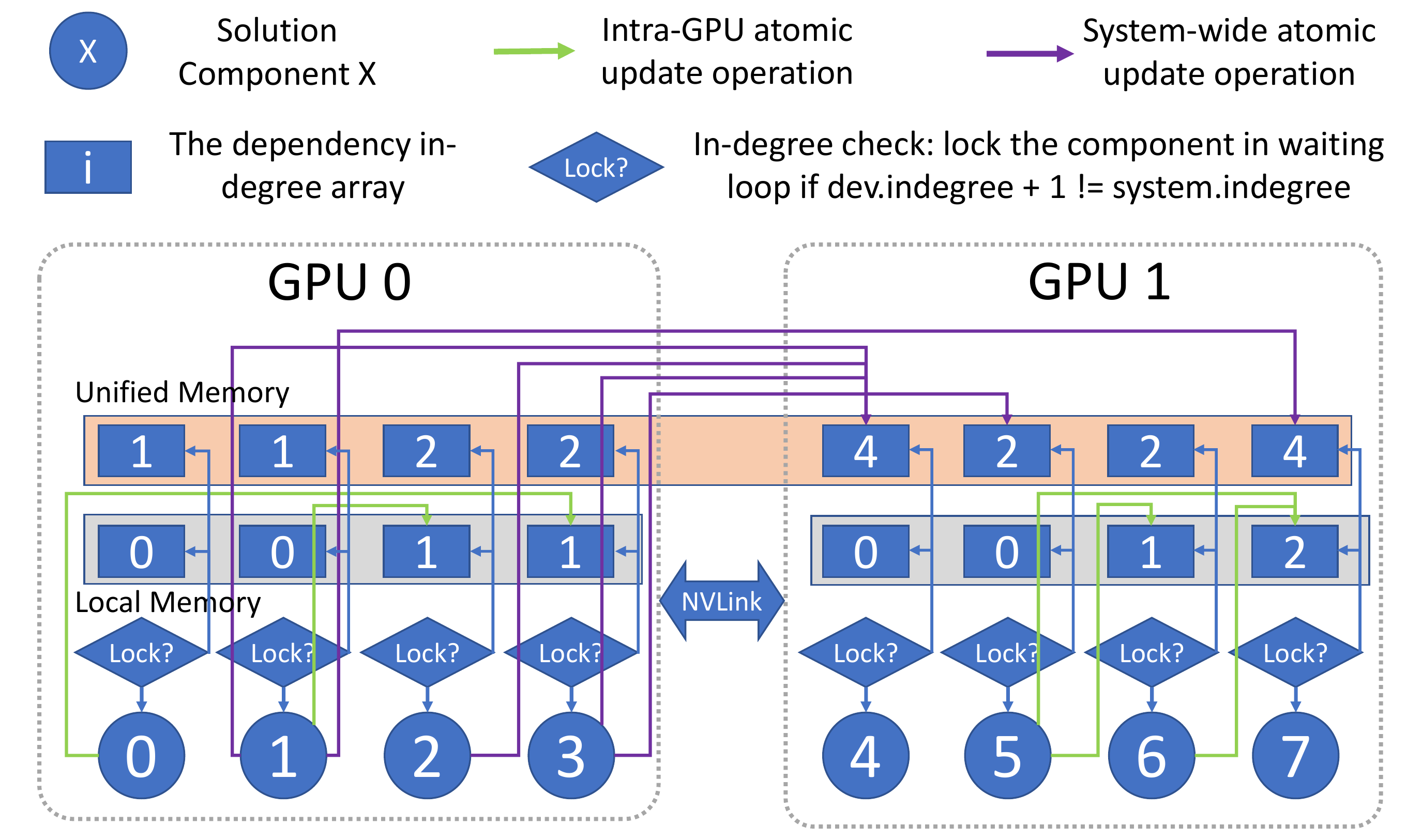}
	
	\end{minipage}
	\caption{SpTRSV on multi-GPUs with Unified Memory. We use system-wide and device-wide atomic update operations for \textit{s.in.degree} and \textit{d.in.degree}, respectively.}
   	\vspace{-0.5cm}
	\label{fig:unified}
\end{figure}

However, after workload profiling, we observe two major disadvantages of this approach due to the invisible data migration of Unified Memory and unidirectional dependency feature of the triangular matrix. 
\Cref{fig:unified} illustrate the data communication model of the Unified Memory approach. In this figure, two GPUs are connected with NVLink. The GPUs communicate the dependency values $in.degree$ using both unified and device memory. The $left.sum$s are also allocated similarly so we do not show them in this figure. 
From the figure, we observe that all $x_1$, $x_2$ and $x_3$ components solved by GPU 0 need to access the $s.in.degree$ of $x_4$ remotely to update the value while the busy-wait loop of $x_4$ on GPU 1 also needs to access the value on unified memory continuously to check if all dependencies have been met. Since the unified memory is designed to manage the data transparently to users, GPU 0 and GPU 1 will compete for the $s.in.degree$ value of $x_4$. As a result, a huge number of page thrashing is generated and the pages containing the data bounce back and forth between GPU 0 and GPU 1.
In addition, we can see that system-wide atomic updates are unidirectional, i.e., the components with larger numbers ($x_{4...7}$) always depend on proceeding components ($x_{0...3}$) in forward substitution). Consequently, GPU 1 has longer waiting time than GPU 0. This can severely limit the scalability of SpTRSV on a multi-GPU system.

\subsection{Characterising the Page Thrashing of Unified Memory}

To evaluate the performance implication of page thrashing on SpTRSV with the Unified Memory approach more extensively, we perform experiments on a \textit{NVIDIA V100-DGX-1} multi-GPU system. 
The DGX-1 system is equipped with eight of the NVIDIA Tesla V100 GPUs. The GPUs are organized in a hybrid cube-mesh interconnection network topology. The twelve edges of the cube are connected through NVLink and two of
the six faces have their diagonals connection\cite{li2018tartan}. 


\begin{figure}[t]
	\centering
	\begin{minipage}[h]{0.235\textwidth}
		
		\includegraphics[width=1\textwidth]{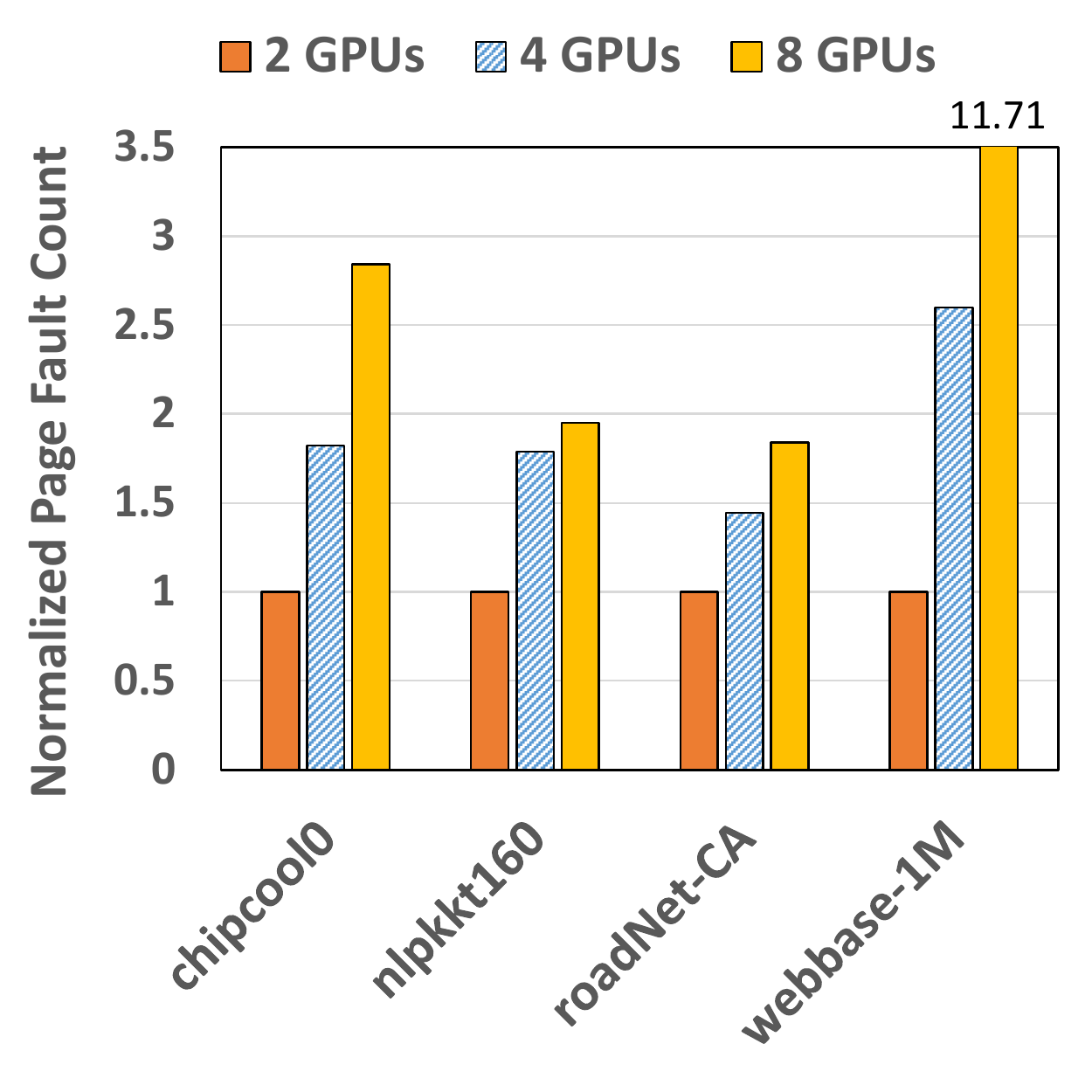}
		\vspace{-0.5cm}
		\subcaption{Page Fault Count}\label{fig:pagefault}
	\end{minipage}
	\begin{minipage}[h]{0.235\textwidth}
		
		\includegraphics[width=1\textwidth]{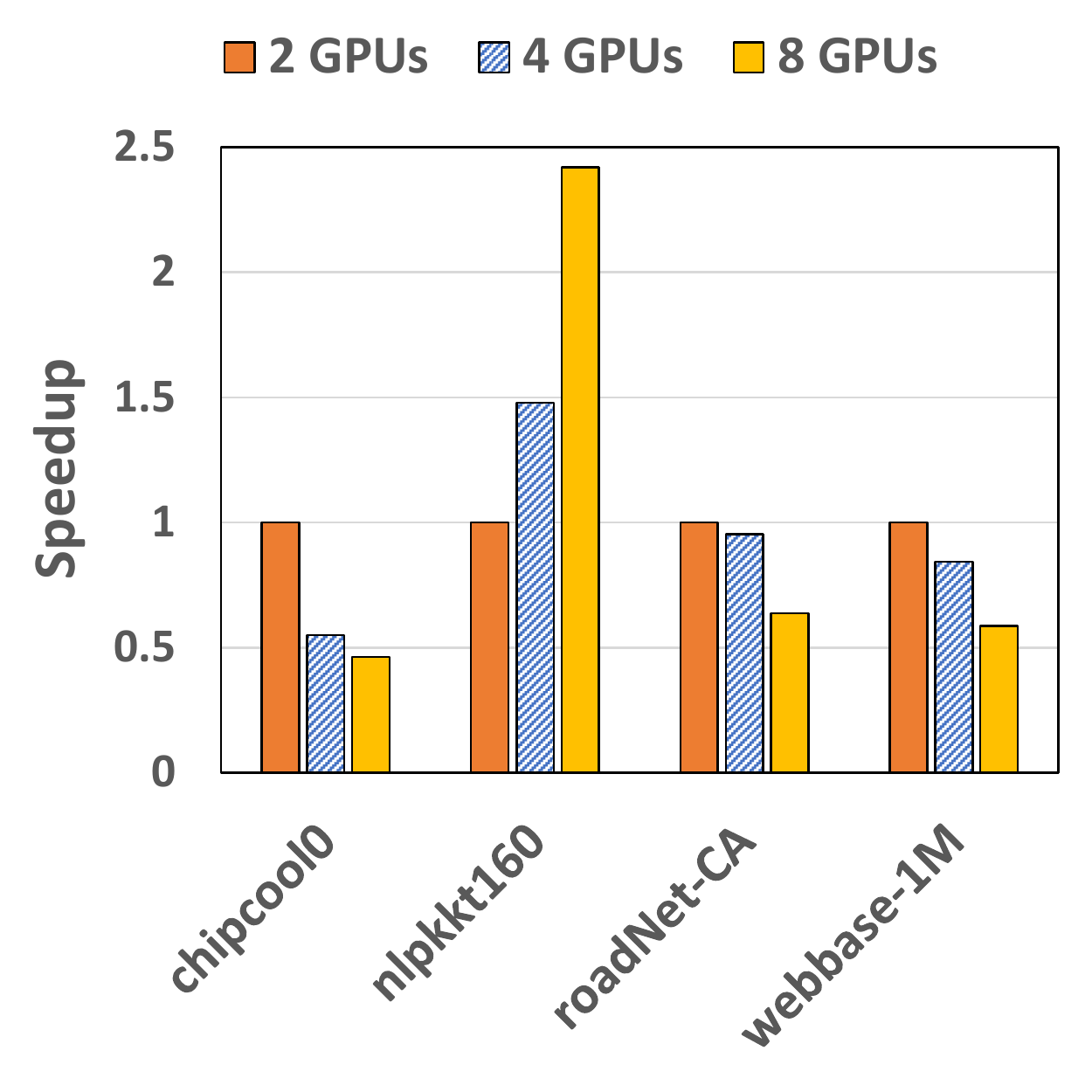}
		\vspace{-0.5cm}
		\subcaption{Normalized Performance}\label{fig:executionUnified}
	\end{minipage}
	\caption{ The effect of page thrashing on SpTRSV due to data contention in the unified memory.}
		\vspace{-0.5cm}
	\label{fig:Unifiedperformance}
\end{figure}

Figure \ref{fig:Unifiedperformance} shows the profiling results of running SpTRSV algorithm on DGX-1 with the Unified Memory approach using four representative matrices from SuiteSparse\cite{sparsematrix}. The results are reported by normalizing to the 2-GPU execution since we want to make observations on multi-GPU systems. 
\Cref{fig:pagefault} demonstrates that the number of page faults caused by the data contention on Unified Memory increases with the growing number of GPUs. This is due to the fact that the memory allocated on the unified memory to compute and track the dependencies (in-degrees) gets updated by system-wide atomic operations (e.g. \textit{s.atomic.decr}) by all GPUs. 
In this case, the required pages containing the shared data bounce back-and-forth 
among different GPUs to update different operations, which in turn have detrimental effect on the execution time of SpTRSV as shown in \Cref{fig:executionUnified}. For example, except for matrix \textit{nlpkkt160} which has the highest potential for parallelism with the least amount of dependencies, other matrices exhibit performance degradation on the increasing number of GPUs (from 2 to 8) even though there are more computing resources allocated progressively on a 8-GPU system.

In summary, due to 
the high amount of page faults caused by the updates from multiple GPUs on shared data in the Unified Memory space, it is not viable to employ Unified Memory in designing multi-GPU SpTRSV.


%% file: sec-shmem.tex
\section{Zero-copy SpTRSV with NVSHMEM}
\label{sec:nvshmem}
	

In this section, we design a novel zero-copy sparse triangular solver on multi-GPU systems that employs NVIDIA's \textit{NVSHMEM} technology based on OpenSHMEM \cite{openshmem_website,potluri2016simplifying,potluri2017openshmem} for efficient inter-communication among GPUs. This approach avoids memory contention among GPUs and eliminates the automatic data migration through the Unified Memory.   

\subsection{NVSHMEM Overview}



NVSHMEM \cite{potluri2016simplifying} extends OpenSHMEM's interface to allocate pinned memory on NVIDIA GPUs that are distributed and interconnected with NVLink or PCIe. Each GPU can be considered as a processing element (PE) and NVSHMEM allocates symmetric heap on each PE. These symmetric heaps together constitute the global address space, hence conforming to the Partitioned Global Address Space (PGAS) programming model. In addition, for better overlapping between communication and computation, NVSHMEM enables GPU-initiated fine-grained point-to-point communication with \emph{get} and \emph{put} operations. NVSHMEM also provides synchronization primitives for data allocation on the heap.


\begin{figure}[t]
	\centering
	\begin{minipage}[h]{0.47\textwidth}
		
		\includegraphics[width=1\textwidth]{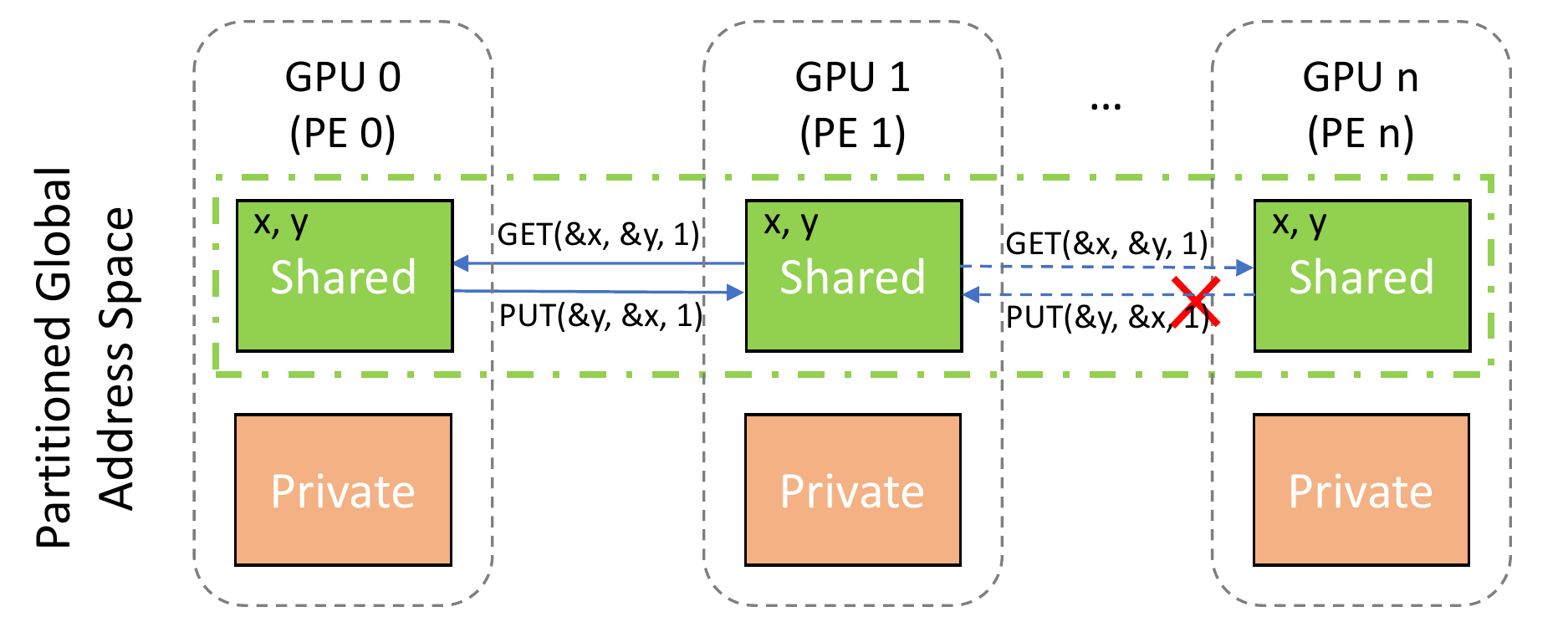}
	\vspace{-0.4cm}
	\end{minipage}
	\caption{NVSHMEM memory space and communication model. To ensure correctness, the shared data cannot be updated simultaneously.}
		\vspace{-0.5cm}
	\label{fig:nvshmem}
\end{figure}

\Cref{fig:nvshmem} shows the memory layout and communication model of NVSHMEM. During initialization, each instance GPU is assigned with a unique ID, called the PE ID. PEs within a NVSHMEM job can share data through the globally accessible memory that is allocated using the NVSHMEM allocation API (e.g. \emph{nvshmem\_malloc()}). Otherwise, memory allocated using any other CUDA methods (e.g. \textit{cudaMalloc()}) is private to the allocating PE and is not accessible by other PEs. Shared memory objects allocated in the global address space (e.g., $x$ and $y$ in \Cref{fig:nvshmem}) are collective and symmetric across all the PEs in a job. This requires all PEs to participate in the allocation call and  pass the same value in the size argument during data allocation. For communication, PEs use \textbf{put/get} API of NVSHMEM to write/read the shared data value to/from remote GPUs. For example, GPU 0 uses command \textit{PUT(\&y, \&x, 1)} in \Cref{fig:nvshmem} to translate the local source pointer to the destination pointer at GPU 1.

\subsection{SpTRSV Design with NVSHMEM}
\textbf{Challenges in accessing shared memory.} 
Although NVSHMEM provides support for fine-grained GPU-initiated communication mechanism, if the intermediate data structures for maintaining dependencies and partial sums of SpTRSV are distributed across GPUs evenly and are allocated on the symmetric heap, accessing and updating such data (dependencies and partial sums) across GPUs can still create contention. 
In this regard, the next challenge in designing SpTRSV is to efficiently update the intermediate arrays (\textit{in.degree} and \textit{left.sum}) without using the system-wide atomic operations. A naive approach is to apply \textit{Get-Update-Put} operations to remotely read the intermediate data from the target PE and then write back the updated value. However, this naive approach introduces data-race hazard when multiple threads compete to update the same value simultaneously. Since NVSHMEM relaxes the total ordering requirement for get operation, to ensure correctness, an \emph{nvshmem\_fence} operation is required for each of the get operations to access the shared data. In addition, to ensure completeness of get/put operations of symmetric data objects as well as to ensure visibility of the updates to all the PEs, SHMEM quiet semantics (with \emph{shmem\_quiet}) needs to be enforced. However, these requirements are too restrictive.
As a result of these restrictions, during the solver phase, only one PE will be able to operate on shared data while others have to wait until the memory is free and visible as \Cref{fig:nvshmem} shows. Similar to the unified memory approach, the naive approach also suffers from data contention that will adversely affect the performance.

\begin{figure}[t]
	\centering
	\begin{minipage}[h]{0.48\textwidth}
		
		\includegraphics[width=1\textwidth]{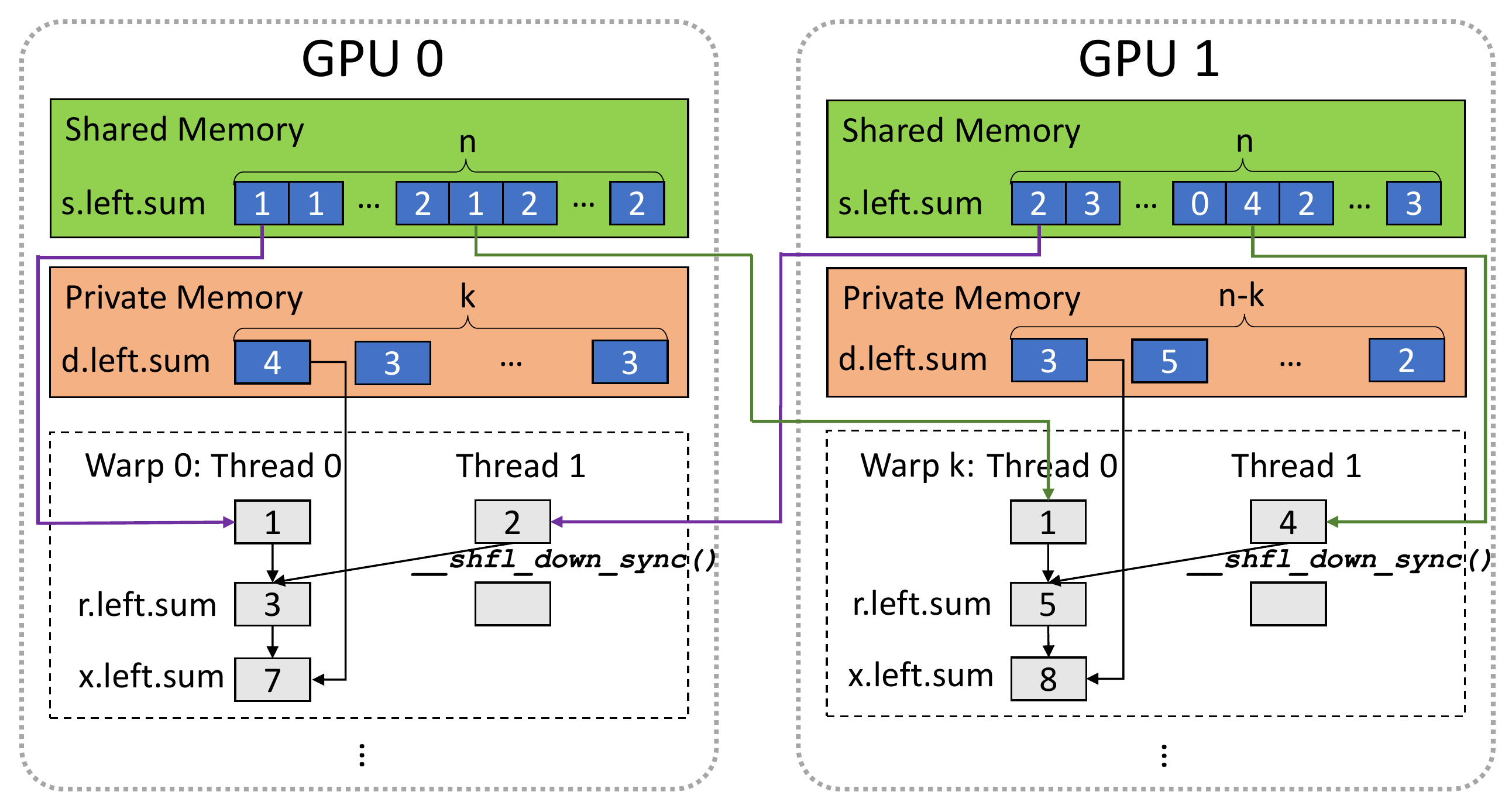}
	
	\end{minipage}
	\caption{The read-only communication model of our SpTRSV with NVSHMEM and the warp-level parallel reduction mechanism. Note that only the $left.sum$ arrays are shown, but the $in.degree$ arrays are also allocated on the symmetric heap.}
		\vspace{-0.5cm}
	\label{fig:Sptrsvnvshmem}
\end{figure}

\textbf{Read-only Inter-GPU Communication.} 
To circumvent these restrictions of the naive approach, we propose a read-only inter-GPU communication model to avoid the data race and update the intermediate arrays asynchronously across all the PEs, as shown in Figure~\ref{fig:Sptrsvnvshmem}. On each PE, we allocate two intermediate arrays of size $n$ on the symmetric heap to hold the system-wide $in.degree$ and $left.sum$ (\textit{s.in.degree} and \textit{s.left.sum}), where $n$ is equal to the number of solutions. Thus, each PE first accumulates updates locally during system-wide intermediate data update. Once local computation is done, the final intermediate value for a component $x_i$ is computed by collecting all system-wide intermediate values of $x_i$ from different GPUs by using the asynchronous \textbf{get} API. We read the values from different PEs into different threads within the same warp to achieve parallel remote memory access and fully utilize the interconnect bandwidth. To sum the values together, we employ the warp-level parallel reduction mechanism to reduce the complexity of the sum-loop operation from O($\#PE$) to O($log(\#PE)$) by using \textit{\_\_shfl\_down\_sync()} operation.

\begin{algorithm} \small
	\caption{SpTRSV on multiple GPUs with NVSHMEM}\label{alg:Shmem}
		\begin{algorithmic}[1]
	    	\Statex $\textbf{Input: } col.ptr_{n} \text{,  } row.idx_{nnz} \text{,  } val_{nnz} \text{,  } B_{n}$
		    \Statex $\textbf{Output: } X_{n}$
			\Procedure{SpTRSV-nvshmem}{}
			\ForAll {$PE$} \Comment{parallel in GPUs}
			\State $dev.x \gets \text{allocate x for device}$
			\State $dev.n \gets \text{size of x in device}$
			\State $\text{cudaMalloc(*d.left.sum, dev.n)}$\label{line:alloc_device_nv_start}
			\State $\text{cudaMalloc(*d.in.degree, dev.n)}$
			\State $\text{cudaMemset(*d.left.sum, 0)}$ 
			\State $\text{cudaMemset(*d.in.degree, 0)}$\label{line:alloc_device_nv_end}
			
			\State $\text{nvshmem\_Malloc(*s.left.sum, n)}$\label{line:nv_alloc_start}
			\State $\text{nvshmem\_Malloc(*s.in.degree, n)}$
			\State $\text{Memset (*s.left.sum, 0)}$
			\State $\text{Memset (*s.in.degree, 0)}$\label{line:nv_alloc_end}
			
		    \LState $\textit{Get s.in.degree for device x:}$\label{line:device_in_deg_start}
			\ForAll {$i \in dev.nnz$} \Comment{parallel in threads}
		    \State	$\text{d.atomic.incr(\&s.in.degree[row.idx[i]]}$
			\EndFor\label{line:device_in_deg_end}
			
			\LState $\textit{Solve device x:}$
			\ForAll {$i \in dev.x$} \Comment{parallel in warps}
			\State $\text{x.in.degree[i]} \gets 0$
			\While {$\text{d.in.degree[i]+1} \neq \text{x.in.degree[i]}$ }\label{line:check_dep_start}
			\ForAll{$d \in ngpu$} \Comment{parallel in threads}
			\If{$\text{r.in.degree[i][d]} \neq 0$}
			\State $\text{r.in.degree[i][d]} \gets \text{get(s.in.degree[i],d)}$
			\EndIf
			\EndFor
			\State $\text{x.in.degree[i]} \gets \text{reduction(*r.in.degree[i])}$
			\EndWhile \label{line:check_dep_end}
			\ForAll{$d \in ngpu$} \Comment{parallel in threads}
			\State $\text{r.left.sum[i][d]} \gets \text{get(s.left.sum[i],d)}$
			\EndFor
			\State $\text{x.left.sum[i]} \gets \text{reduction(*r.left.sum[i])}$
			
			\State $\text{x[i]} \gets \text{b[i] - d.left.sum[i] - x.left.sum[i]}$
			\State $\text{x[i]} \gets \text{x[i]/val[col.ptr[i]]}$
			
			\ForAll {$j \in nnz_i $} \Comment{parallel in threads}
			\State $ \text{rid} \gets \text{row.idx[j]}$
			\If {$rid \in dev.x$}
			\State $\text{d.atomic.add(\&d.left.sum[rid], val[j]x[i])}$
			\State $\text{d.atomic.incr(\&d.in.degree[rid])}$
			\Else
			\State $\text{d.atomic.add(\&s.left.sum[rid], val[j]x[i]))}$
			\State $\text{d.atomic.decr(\&s.in.degree[rid])}$
			\EndIf
			\EndFor
			
			\EndFor
			
			\EndFor
			
			\EndProcedure
		\end{algorithmic}	
\end{algorithm}

Algorithm \ref{alg:Shmem} shows the pseudocode of our SpTRSV design with NVSHMEM. In this design, every GPU in the system is considered as one PE and executes the same program to solve the linear system. The algorithm maintains two device-wide intermediate arrays ($d.in.degree$ and $d.left.sum$) on the private memory for fast updating intermediate values within the same thread block (\linesref{line:alloc_device_nv_start}{line:alloc_device_nv_end}). The algorithm also initializes the system-wide intermediate arrays using \textit{nvshmem\_malloc()} API (\linesref{line:nv_alloc_start}{line:nv_alloc_end}). To indicate how many warps have to be finished in advance before each component can proceed to the solver step, the algorithm calculates the system-wide $in.degree$ similar to the unified memory approach. However, the difference is that it processes partial $in.degree$ values locally for each PE so that no inter-GPU communication is required during this step (\linesref{line:device_in_deg_start}{line:device_in_deg_end}). 

After calculating the partial $in.degree$ values, the algorithm proceeds to solve each $x_i$ using two phases: \emph{lock-wait} and \emph{solve-update}, which is similar to the Unified Memory implementation. For the \emph{lock-wait} phase, the algorithm employs the read-only inter-GPU communication model to collect all system-wide $in.degree$ values within the \textit{while} loop to check if all dependencies have finished (\linesref{line:check_dep_start}{line:check_dep_end}). Even though the remote memory access will increase certain latency to the kernel, the memory latency can be hidden by processing numbers of warps simultaneously in each GPU. {Additionally, to lower the bandwidth requirement, we check the temp indegree value ($r.in.degree$) before each remote read. If $r.in.degree = 0$, it indicates that the remote dependencies are satisfied. Then we pass the remote access to the node to reduce the number of interconnect communication. }
Then, for \emph{solve-update} phase, we use similar method to collect all the system-wide $left.sum$ to solve the component $x$ and update the intermediate data locally for its dependents using hybrid memory system(line 28-35). Note that this method still employs device-wide atomic operations to update the intermediate value as multiple updates from different warps of one PE may happen simultaneously.  

By employing the read-only inter-GPU communication model to design SpTRSV with NVSHMEM on a multi-GPU platform, we isolate the execution for each PE and avoid page migration cost from data contention and system-wide synchronization compared to the Unified Memory approach. 

%% file: sec-taskmodel.tex
\section{Fine-Tuning Workload Dependency}
\label{sec:task_model}
In the previous section, we have discussed the general strategies for reducing contention and exchanging data in SpTRSV, particularly allocating shared data on the symmetric heap and relying on the one-sided communication primitives in NVSHMEM for inter-GPU communication. However, the performance of SpTRSV is still restricted by execution time imbalance arising from dependency disparity across different components (i.e., $x_i$). 



		
	

Static distribution of workload forces GPUs with larger IDs to always wait longer than the smaller ID ones, since the former can only resume work when all the dependencies from GPUs with smaller ID have been met. As we mentioned in \Cref{sec:back}, because the baseline workload distribution model of multi-GPU system partitions components $x$, columns of $L$ matrix and right-hand-side $b$ and allocates the partitions to each GPUs in the ascending order of the component number, the dependencies among the GPUs are essentially unidirectional. This may negate execution time balance and performance of SpTRSV on multi-GPU systems. 


\begin{figure}[t]
	\centering
	\begin{minipage}[h]{0.43\textwidth}
		
		\includegraphics[width=1\textwidth]{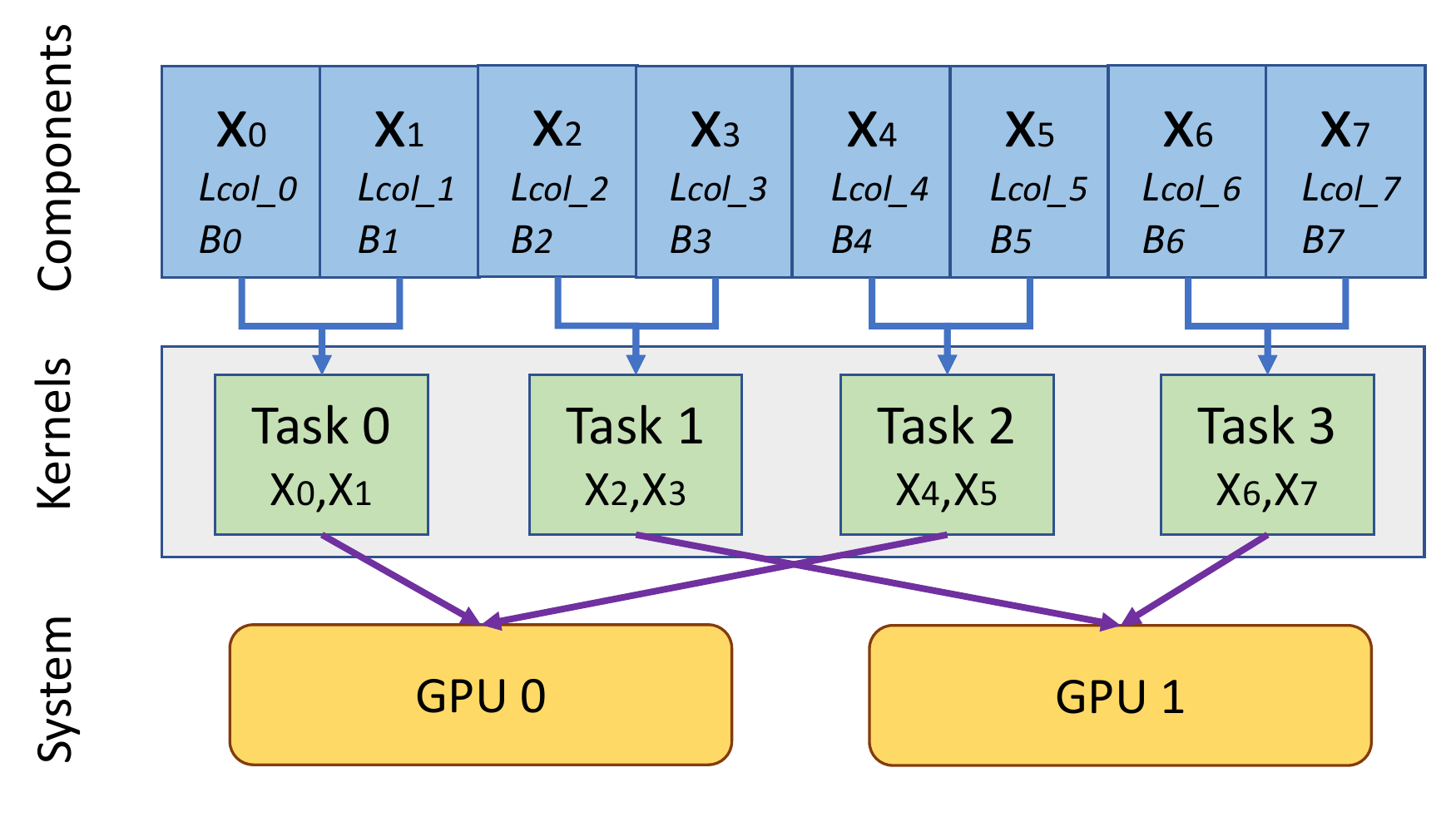}
	\end{minipage}
		\vspace{-0.2cm}
	\caption{Grouping components in tasks for effective workload distribution.}
		\vspace{-0.5cm}
	\label{fig:tranfer}
	
\end{figure}

\begin{figure*}[t]
	\centering
	\begin{minipage}[h]{0.94\textwidth}
		
		\includegraphics[width=1\textwidth]{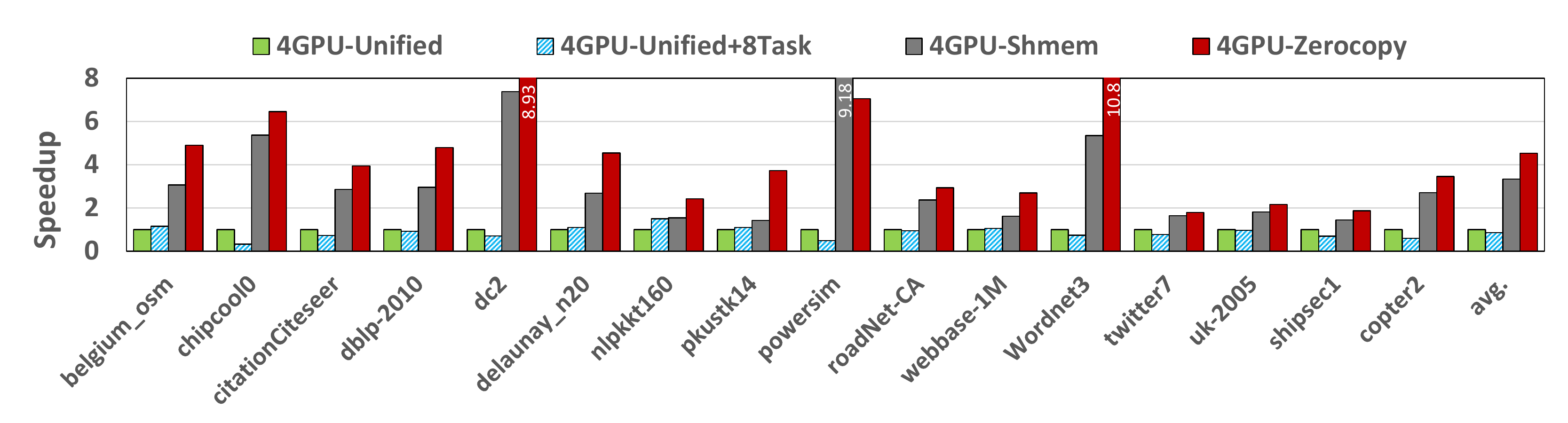}
	
	\end{minipage}
		\vspace{-0.2cm}
	\caption{The speedup of SpTRSV under different design scenarios on 4 GPUs DGX-1 system.}
		\vspace{-0.2cm}
	\label{fig:overallperformance}
\end{figure*}

To tackle the unidirectional dependency issue, we propose a 
fine-tuned task distribution module to utilize the spatial locality of dependent components in SpTRSV on multi-GPU systems. As shown in \cref{fig:tranfer}, we first divide total components into small \emph{component-tasks} and group them together based on a user-specified parameter for  component count in a component task. Each task has the same number of components $x$, columns of the $L$ matrix and right-hand-side $b$. This process is done during the data distribution. Each task is considered as the smallest unit of workload to schedule. Tasks are allocated to the GPUs in a round-robin order based on the available memory of the GPUs. During the execution of SpTRSV, each task is launched as one GPU kernel and the components within the task are assigned to thread warps for parallel execution. Note that, all tasks scheduled on the same GPU share same sets of intermediate arrays in the NVSHMEM global address space. Atomic operations ensure proper updates of the intermediate arrays by different tasks on the same GPU. 


Our task distribution module enables scheduling the smaller ID components aggressively so that the waiting time before starting the solvers for larger ID components is reduced. 
However, finer granularity of tasks lower the amount of components in each kernels and may incur more overhead in kernel launching. In this case, determining optimal number of tasks becomes a new trade-off between fine-grained task scheduling and long kernel scheduling overhead. In our work, we conduct a sensitivity study about this trade-off and we observe that even with a small number of tasks per GPU (e.g. 4 tasks per GPU), our SpTRSV algorithm is able to balance the execution time of each GPU on a multi-GPU system. We present our experimental results in \Cref{sec:evaluation}.

%% file: sec-evaluation.tex
\section{Evaluation}\label{sec:evaluation}
In this section, we report our experimental results and provide detailed analysis on our proposed SpTRSV design for multi-GPUs system. These include overall performance, sensitivity study for determining the optimal number of tasks to schedule per GPU and scalability of our algorithm with increasing number of GPUs (strong scaling).
\subsection{Experimental Setup}


\textbf{Platform.} We implemented our NVSHMEM-based zero-copy SpTRSV with task model using CUDA 10.1 and test it on the NVIDIA V100-DGX-1 and NVIDIA V100-DGX-2 multi-GPU systems. The DGX-1 system is equipped with eight NVIDIA Tesla V100 GPUs with 16GB memory, where GPUs are inter-connected via NVLinks with 64GB/s  maximum inter-communication bandwidth. The DGX-2 system is equipped with 16 V100 GPUs where GPUs are all-to-all connected through NVSwitch with around 100GB/s bandwidth per node. The two multi-GPUs system is operated by two 20-Core Intel Xeon E5-2698 v4 CPUs with 512 GB memory. Since NVSHMEM communication only can be initiated between P2P-connected GPUs currently, we perform NVSHMEM implementations on up to 4 GPUs on the DGX-1 system that are fully connected \cite{li2018tartan}. 
We used the native \emph{mpirun} job launcher of NVSHMEM --- \textit{hydra}, to launch MPI processes on each GPU node and execute the GPU kernels for the proposed zero-copy SpTRSV. 

\begin{scriptsize}
	\begin{table}[t]\footnotesize
	    \captionsetup{labelfont = bf}
		\centering
		\caption{Test Matrices. "Parallelism" refers to the average available concurrency component per level shown in Figure~\ref{fig:level-sets}}
		\begin{tabular}{|c|r|r|r|r|p{0.48\textwidth}|}
			\hline
			\bf Name &\bf \#Rows & \bf \#Non-Zeros & \bf \#Levels & \bf Parallelism \\
			\hline
			\hline
			belgium\_osm & 1,441,295 & 2,991,265 & 631 & 2,284 \\
			\hline
			chipcool0 & 20,082 & 150,616 & 534 & 38 \\
			\hline
			citationCiteseer & 268,495 & 1,425,142 & 102 & 2,632 \\
			\hline
			dblp-2010 & 326,186 & 1,133,886 & 1,562 & 209 \\
			\hline
			dc2& 116,835 & 441,781 & 14 & 8,345 \\
			\hline
			delaunay\_n20& 1,048,576 & 4,194,262 & 788 &  1,331\\
			\hline
		    nlpkkt160 & 8,345,600 & 118,931,856 & 2 & 4,172,800 \\
		    \hline
		    pkustk14 & 151,926 & 7,494,215 & 1,075 & 141\\
			\hline
			powersim & 15,838 & 40,673 & 24 &  660 \\
			\hline
			roadNet-CA & 1,971,281 & 4,737,888 & 364 & 5,416\\
			\hline
			webbase-1M & 1,000,005 & 2,348,442 & 512 & 1,953 \\	
			\hline
			Wordnet3 & 82,670 & 176,821 & 37 & 2,234\\
			\hline
			shipsec1 & 7,813,404 & 140,874 & 2100  & 67\\
			\hline
			copter2 & 759,952 & 55,476 & 190  & 291\\
			\hline
			twitter7 & 41,652,230 & 475,658,233 & 18,116 & 2,299 \\
			\hline
			uk-2005 & 39,459,925 & 473,261,087 & 2,838 & 1,390,413 \\
			\hline
			
		\end{tabular}
			\vspace{-0.2cm}
		\label{table:benchmarks}
	\end{table}
\end{scriptsize}

\textbf{Test Matrices.} \Cref{table:benchmarks} lists the 14 sparse matrices we selected for our experiments. These metrics have been collected from SuiteSparse \cite{sparsematrix} matrix collection and have been used in previous studies of sparse matrix computations \cite{naumov2011parallel,wang2018,totoni2014structure,anzt2015iterative,Sao2019,park2014Sparsifying,kabir2015stsk,bradley2016hybrid,dufrechou2018level}. We factorized these matrices to generate the sparse L decomposition using MA48 \cite{ma48} from the Harwell Subroutine Library (HSL) \cite{HSL} following the approach of the previous work \cite{liu2017fast}. These matrices cover a variety of application scenarios, as well as a wide range for the size of the components and the number of level-sets as listed in Table~\ref{table:benchmarks}. Two of these matrices, \textit{twitter7} and \textit{uk-2005}, are out-of-memory matrices which have original 21.6GB and 16.8GB input files, respectively. In our design, we first decompose the original matrices and then execute SpTRSV on the generated lower triangular matrices. On average, the intermediate arrays consume 10\% of total memory requirement across all matrices in our experiment. 
For each benchmark, we ran SpTRSV 100 times and report the average execution time.

\subsection{Performance Evaluation}

\begin{figure*}[t]
	\centering
	\begin{minipage}[h]{0.94\textwidth}
		
		\includegraphics[width=1\textwidth]{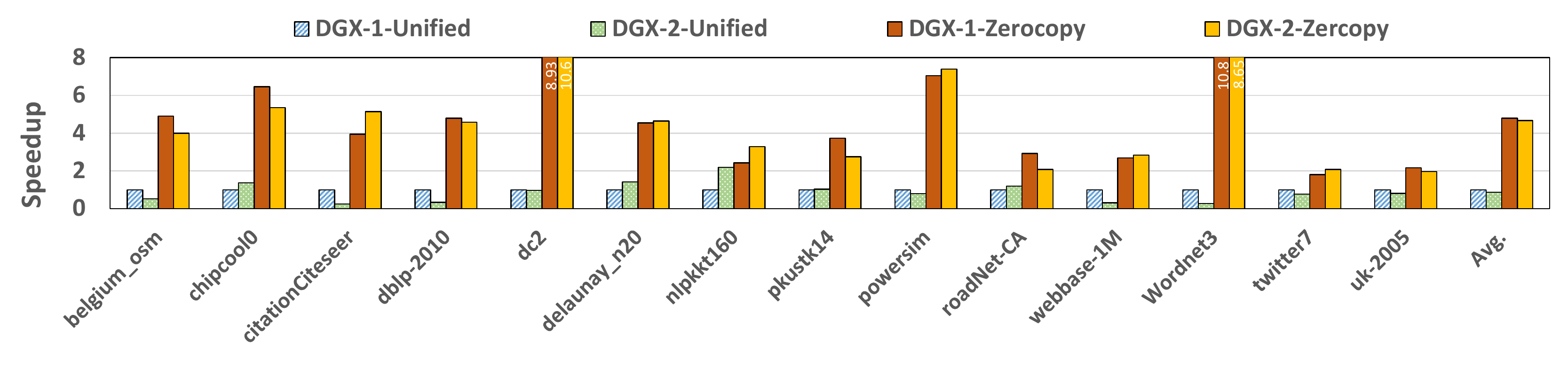}
	
	\end{minipage}
		\vspace{-0.4cm}
	\caption{The speedup of SpTRSV on 4 GPUs DGX-1 and DGX-2 system.}
		\vspace{-0.4cm}
	\label{fig:overallperformance2}
\end{figure*}

\begin{figure*}[t]
	\begin{center}
		\includegraphics[width=0.94\textwidth]{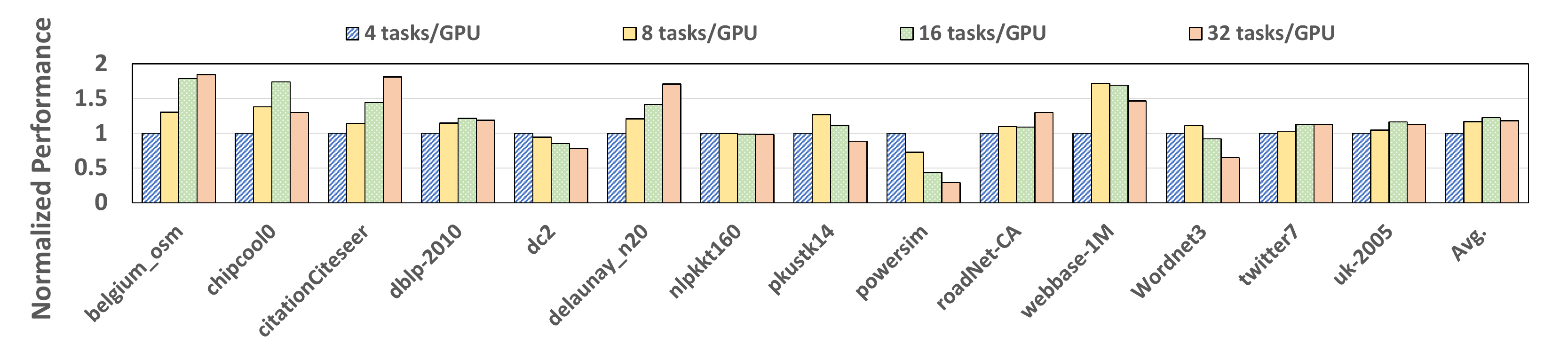}
			\vspace{-0.2cm}
	\caption{The normalized performance of SpTRSV with different number of tasks per GPU on 4 GPUs DGX-1 system.}
	\vspace{-0.4cm}
	\label{fig:sens}
	\end{center}
\end{figure*}

To evaluate the effectiveness of our proposed SpTRSV design, we first compare the performance on a 4-GPU DGX-1 system under several design scenarios as shown in \Cref{fig:overallperformance}: (i) \textit{4GPU-Unified} -- the synchronization-free SpTRSV with Unified Memory (Section 3); (ii) \textit{4GPU-Unified+8task} -- the task-model enabled SpTRSV-Unified with 8 tasks per GPU; 
(iii) \textit{4GPU-Shmem} - the NVSHMEM-based SpTRSV with continued component distribution (Section 4); and (iv) \textit{4GPU-Zerocopy} - the proposed task-model enabled zero-copy SpTRSV with 8 tasks per GPU (Section 5).  
We sum up the execution time of the analysis phase and the solver phase, and normalize the total execution time in each case with respect to the execution time of 4GPU-Unified. From the figure, we obtain several observations.

First, directly imposing the task model on unified memory exacerbates the page fault thrashing problem. Compared to \emph{4GPU-Unified}, the performance of \emph{4GPU-Unified+8task} reduces by $\approx$11\% on average. This is because with finer granularity of tasks, the number of page faults and data contention increase on multi-GPU system.
Second, we observe that \emph{4GPU-Shmem} outperforms \emph{4GPU-Unified} by about 2.33$\times$ on average (up to 8.1$\times$). Using NVSHMEM and the read-only communication model, \emph{4GPU-Shmem} successfully avoids the page-fault and page thrashing across the mutli-GPUs. 
Finally, the proposed design scenario, \emph{4GPU-Zerocopy}, improves the overall performance by 3.53$\times$ (up to 9.86$\times$) over \emph{4GPU-Unified}. In addition, we observe that \emph{4GPU-Zerocopy} achieves even better performance for the matrices with higher degree of \textit{parallelism} (e.g. \textit{dc2}, \textit{nlpkkt160}, \textit{powersim} and \textit{Wordnet3}). This is because with our task model, spatial locality is exploited among dependent components dispatched to the same GPU, achieving better workload balance than \emph{4GPU-Shmem}. 

Next, we conduct the same experiment on DGX-2 systems where GPUs are connected with different network topologies. \Cref{fig:overallperformance2} shows the performance of multi-GPU SpTRSV on DGX-1 and DGX-2 with 4 GPUs and 8 tasks per GPU. We normalize the result in each case to DGX-1-Unified. From this figure, we observe that the proposed Zero-copy SpTRSV can achieve similar performance improvement (3.53$\times$ for DGX-1 and 3.66$\times$ for DGX-2) under both systems. Although the DGX-2 system has higher available inter-connection bandwidth than DGX-1, the algorithm achieves similar speedup in both cases. This indicates that our proposed algorithm can effectively overlap communication in the \textit{Lock-Wait} phase with the computation in the \textit{solve-update} phase. In this case, the algorithm fully leverages the bandwidth of the communication channels as well as the computing resources of all the GPUs.

\subsection{Sensitivity Study}

As mentioned in \Cref{sec:task_model}, the number of tasks per GPU can affect the efficiency of NVSHMEM-based SpTRSV on a multi-GPU system. To understand how the task model impacts the design choice, we examine the workload balance and performance gain of zero-copy SpTRSV under a variety of choices on the number of tasks per GPU. 

Figure~\ref{fig:sens} shows the average performance gain across all benchmarks with different number of tasks when executing zero-copy SpTRSV using 4 GPUs. The performance is normalized to the result of the 4 tasks per GPU scenarios. From the figure, we can observe that due to a better workload balance, finer granularity of tasks often contributes to the higher performance improvement. Specifically, comparing with 4 tasks per GPU, allocating 16 tasks per GPU can bring on average 22\% and up to 78\% performance enhancement.

Nevertheless, increasing the tasks number per GPU is not always beneficial for some matrices. For instance, ``\textit{webbase-1M}" attains its best performance (i.e., 69\% over 4 tasks) when there are 8 tasks per GPU. With extra tasks, the performance starts to degrade. This is essentially a trade-off: more tasks per GPU leads to finer-grained communication and better workload balance, but at the same time, suffer from higher scheduling overhead to issue tasks to different GPUs. 

\subsection{Scalability Study}


\begin{figure}[t]
	\centering
	\begin{minipage}[h]{0.24\textwidth}
		
		\includegraphics[width=1\textwidth]{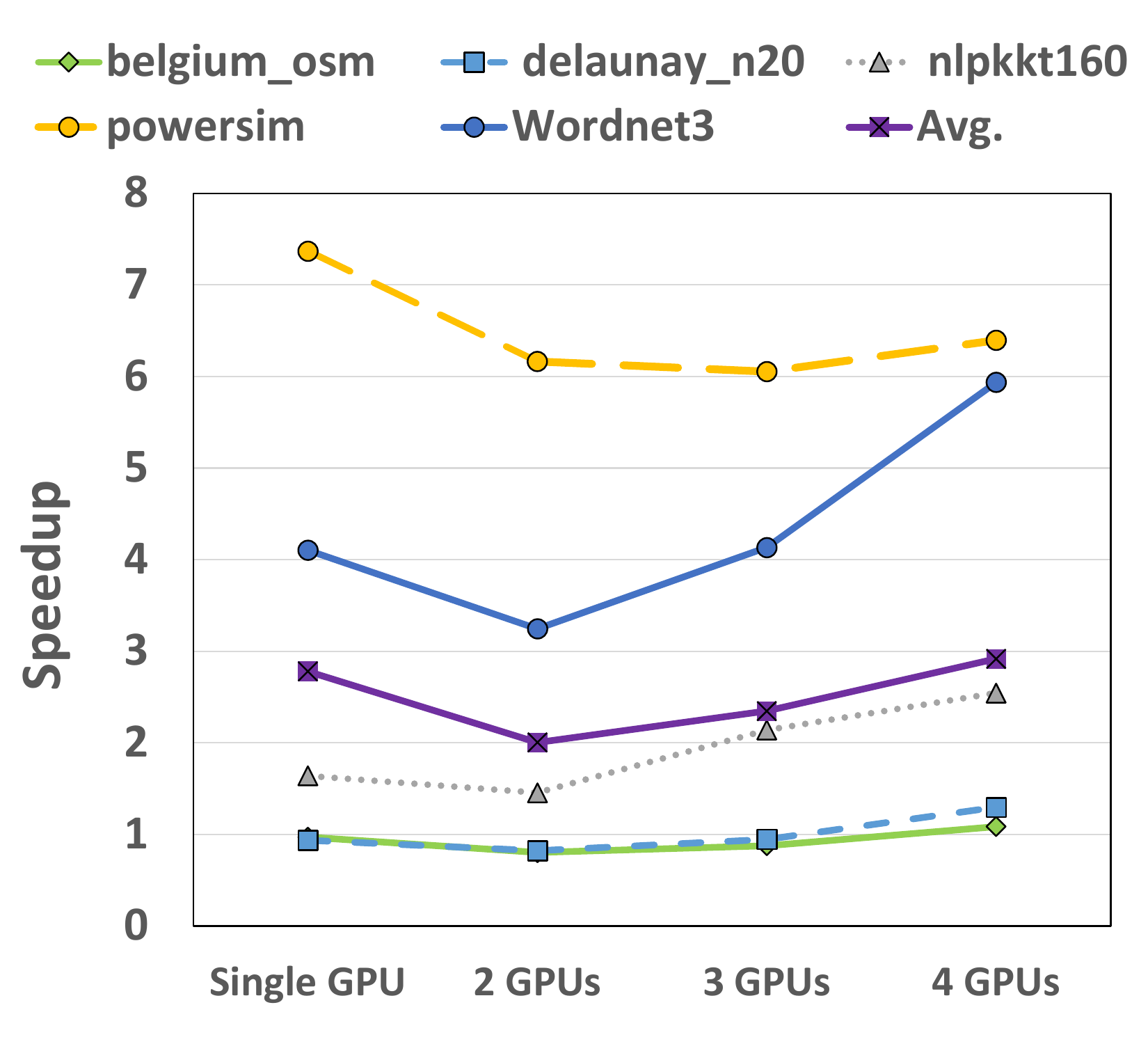}
		\vspace{-0.5cm}
		\subcaption{Scalability in DGX-1}\label{fig:scale-dgx1}
	\end{minipage}
	\begin{minipage}[h]{0.24\textwidth}
		
		\includegraphics[width=1\textwidth]{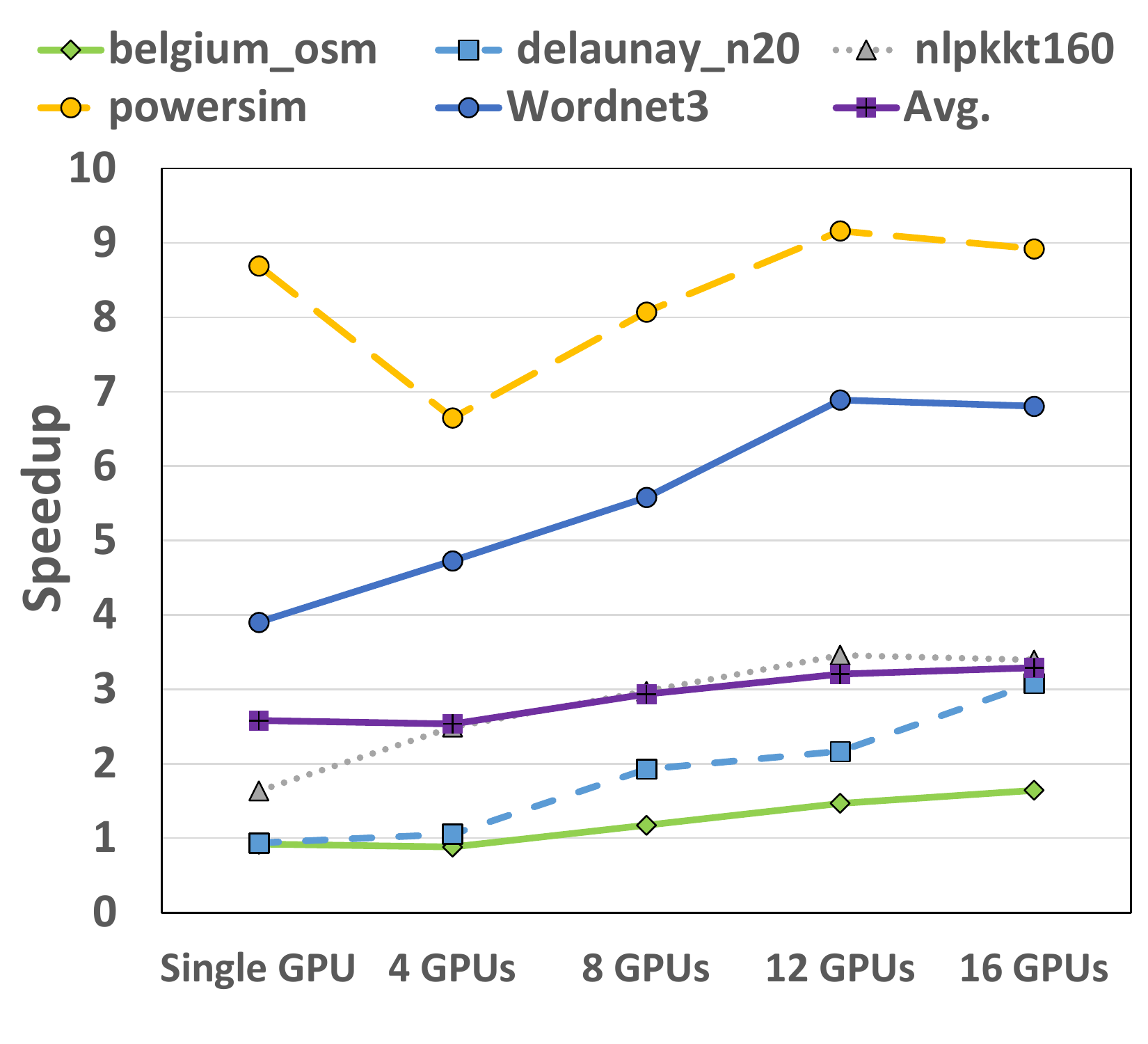}
		\vspace{-0.5cm}
		\subcaption{Scalability in DGX-2}\label{fig:scale-dgx2}
	\end{minipage}
	\caption{Normalized speedup of SpTRSV in DGX-1 and DGX-2 over \textit{cusparse\_csrsv2()}.}
		\vspace{-0.4cm}
	\label{fig:scale}

\end{figure}

We also evaluate the scalability of our design with increased number of GPUs. For the DGX-1 system, since the current NVSHMEM only supports GPUs directly connected (P2P) with NVLinks, we can only implement our NVSHMEM on up to 4 GPUs. For the DGX-2 system, we scale up to 16 GPUs since the GPUs in DGX-2 are all-to-all connected with NVSwitch. We fix the total number of tasks to 32 and normalize the execution time to the single-GPU \textit{csrsv2()} kernel from NVIDIA's \emph{cuSPARSE} library\cite{cusparse}. 

Figure \ref{fig:scale-dgx1} compares the scalability of our SpTRSV in DGX-1 system with up to 4 GPUs. In addition to the average speedups, we also show the 5 matrices that exhibit distinct characteristics. As can be seen, the proposed zero-copy SpTRSV outperforms the \textit{cusparse\_csrsv2} thanks to the fine-grained communication, synchronization-free execution\cite{liu2017fast}, and orchestrated load balancing design. On average, zero-copy SpTRSV presents scalable performance improvement, with 34\% and 91\% speedups using 4 GPUs over 2-GPU and 3-GPU execution, respectively. This is mainly because involving more GPUs in the DGX-1 system increase the computing resources as well as the active communication bandwidth per GPU.

We also observe that the performance of the single-GPU execution is mostly superior than the 2-GPU and 3-GPU execution due to high-speed on-broad communication while some matrices (e.g. \textit{Wordnet3} and \textit{nlpkkt160}) exhibit superior performance improvement (e.g., 2.69$\times$ from 2 to 4 GPUs). 
To explore which types of matrices can take better advantage from the increasing number of GPUs, we define a \emph{dependency} metric as the average non-zero values per component ($ dependency = NNZ / nRow$) and a \emph{parallelism} metric as the average available components per level ($ parallelism = nRow / nLevel$) for each matrix (as shown in \cref{table:benchmarks}). We find that matrices with lower dependency as well as high parallelism can get more benefits from scaling up the number of GPUs.

\Cref{fig:scale-dgx2} shows the strong scaling results of the zero-copy SpTRSV on the DGX-2 system with up to 16 GPUs. We observe that the average performance improvement from the increasing number of GPUs for DGX-2 is flatter compared to the scaling results on the DGX-1 system.
For such a system architecture where all GPUs are P2P-connected through switches, we observe the active bandwidth per GPU will maintain constant with more GPUs involved. 

In summary, 
the scalability of SpTRSV (e.g., in future systems with a larger number of GPUs) not only depends on the dependency and parallelism metrics for a sparse matrix, but also on the intra-node network design and the signaling technologies.


%% file: sec-conclusion.tex
\section{Related Work}

\textbf{Dependency Elimination for SpTRSV.} Concurrent data structures are fundamental building-blocks for real-world applications. Existing works have proposed various novel data structures to handle the dependencies inside SpTRSV \cite{vuduc2002automatic,vuduc2003automatic,mayer2009parallel, wolf2010factors,wang2018,Sao2019,EfficientLu2020}. For better reusing the right-side-hands on \emph{Sunway} architecture, Wang et al. \cite{wang2018} tile the sparse matrix to control the data flow and explore inter-level parallel for SpTRSV. Ramakrishnan et al. \cite{Sao2019} propose a 3D sparse structure to replicate the dependent data for avoiding expensive communication. Comparatively, our framework load from the raw CSC data directly, avoiding unnecessary data-format conversion to a specific data-format. Additionally, our task-pool based inter-GPU communication mechanism is perpendicular to these existing techniques.

\textbf{One-Sided Communication for SpTRSV.} Previous work \cite{ding2020leveraging} transmits the dependency values of SpTRSV asynchronously between MPI ranks through one-sided broadcast. Compared to this work, our proposal aims for enabling fine-grained P2P communication across multi-GPUs by exploring the OpenShmem like NVSHMEM which is different from achieving data passing through CPU-based methodology such as MPI. Furthermore, our method achieves synchronization-free value updating by proposing read-only communication model without pre-analyzing the critical path for matrices.

\textbf{Communication in Multi-GPU System.} There have been multiple works discussing multi-GPU communication technologies \cite{arunkumar2017mcm,milic2017beyond,potluri2016simplifying,umhint,orr2017gravel,daoud2016gpurdma}. Some of them \cite{arunkumar2017mcm,milic2017beyond} conduct architectural optimizations to reduce inter-GPU memory traffic for GPGPU applications, while others present new software API for speeding up remote memory access \cite{potluri2016simplifying,umhint}, including the hint API for avoiding page-fault when using the unified memory. However, for SpTRSV, the inner dependency and structural information is data-dependent, and are only known at runtime. Thus, it is not feasible to provide such a hint for pinning specific data in a specific region at development time. As a comparison, our zero-copy SpTRSV approach leverages the latest NVSHMEM library for fine-grained communication and efficient memory sharing, achieving considerable performance advantages. 

In addition, targeting multi-node multi-GPU systems, several works \cite{orr2017gravel,daoud2016gpurdma} leverage CPU to build point-to-point communication for multi-GPU. For example, in the software level, Gravel\cite{orr2017gravel} proposes message queue to pass data to target GPUs while in the hardware-level, GPUrdma\cite{daoud2016gpurdma} builds a direct access channel between GPU and main memory to share data. Comparing to these works, our work targets single-node multi-GPU platform where the communication channel is directly built among the GPUs.
We efficiently leverage the fast inter-GPUs interconnect (e.g., NVlink and NVSwitch) and the latest NVSHMEM library for achieving fine-grained communication towards the complex dependencies of SpTRSV.


\section{Conclusion}

In this work, we propose a multi-GPU zero-copy SpTRSV algorithm to effectively handle the inherent dependencies of sparse triangular solver. Employing NVIDIA's latest NVSHMEM technology as the channel to convey fine-grained dependency information among GPUs, we construct a read-only inter-GPU communication model to avoid severe interconnect and memory contention. Additionally, a task-pool execution model is further proposed to balance the workload among GPUs while overcoming the unidirectional dependency challenge. Evaluations on state-of-the-art NVIDIA V100-DGX-1 and DGX-2 platforms demonstrate that our optimized multi-GPU SpTRSV algorithm can achieve on average 3.79$\times$ and 3.66$\times$ speedup on DGX-1 and DGX-2, with respect to the Unified-Memory design when using 4 GPUs, showing great performance scalability.